\documentclass{PoS}
\usepackage{amsmath}
\usepackage{subfigure}
\usepackage{float}
\usepackage{multirow}

\newcommand{\cyp}{Department of Physics, University of Cyprus, P.O. Box 20537, 1678 Nicosia, Cyprus}
\newcommand{\cyi}{ Computation-based Science and Technology Research Center, The Cyprus Institute, 20 Kavafi Str., Nicosia 2121, Cyprus}
\newcommand{\poz}{ Faculty of Physics, Adam Mickiewicz University, Umultowska 85, 61-614 Pozna\'{n}, Poland}
\newcommand{\teml}{Temple University,1925 N. 12th Street, Philadelphia, PA 19122-1801, USA}

\newcommand{\PKU}{
	School of Physics, Peking University, Beijing 100871, China}
\newcommand{\CICQ}{
	Collaborative Innovation Center of Quantum Matter, Beijing 100871, China}
\newcommand{\CHEP}{
	Center for High Energy Physics, Peking University, Beijing 100871, China}
\newcommand{\LNPT}{
	State Key Laboratory of Nuclear Physics and Technology, Peking University, Beijing 100871, China}
\newcommand{\desy}{
	NIC, DESY, Platanenallee 6, D-15738 Zeuthen, Germany}
\newcommand{\wup}{Department of Physics, Bergische Universit\"{}t Wuppertal, Gaussstr. 20, 42119 Wuppertal, Germany}
\newcommand{\bonn}{Institut f\"{u}r Strahlen- und Kernphysik, Rheinische Friedrich-Wilhelms-Universit\"{a}t Bonn,  Nussallee 14-16, 53115 Bonn}
\renewcommand{\figurename}{Fig.}

\title{Parton distribution functions of $\Delta^+$ on the lattice}

\ShortTitle{Parton distribution functions of $\Delta^+$ on the lattice}

\author{Yahui Chai$^a$, Yuan Li$^a$, \speaker{Shicheng Xia} $^a$, Constantia Alexandrou$^{b,e}$, Krzysztof Cichy$^c$, Martha Constantinou$^d$, Xu Feng$^{a,f,g,h}$, Kyriakos Hadjiyiannakou$^b$, Karl Jansen$^i$, Giannis Koutsou$^b$, Chuan Liu$^{a,f,g}$, Aurora Scapellato$^{e,j}$, Fernanda Steffens$^k$
\\
\\
\llap{$^a$} \PKU \\
\llap{$^b$} \cyi \\
\llap{$^c$} \poz \\
\llap{$^d$} \teml \\
\llap{$^e$} \cyp \\
\llap{$^f$} \CICQ \\
\llap{$^g$} \CHEP \\
\llap{$^h$} \LNPT \\
\llap{$^i$} \desy \\
\llap{$^j$} \wup \\
\llap{$^k$} \bonn

}

\abstract{
   	We present results for renormalized matrix elements related to the unpolarized quasi-distribution function of the $\Delta^+$ baryon making use of the large momentum effective theory. Two ensembles of $N_f=2+1+1$ twisted mass fermions with a clover term and pion masses of 250 MeV and 330 MeV are analyzed. We employ  momentum smearing to improve the overlap with the boosted $\Delta$ state  significantly  reducing in this way 
the statistical error of both two- and three-point functions.  }

\FullConference{The 37th International Symposium on Lattice Field Theory - LATTICE2019\\
	16-22 July, 2019\\
	Central China Normal University, Hilton Hotel Wuhan Riverside, Wuhan, China.}

\begin{document}
\section{Introduction}

Quantum Chromodynamics (QCD), the fundamental theory which describes the strong interaction among
quarks and gluons, can be solved non-perturbatively by numerical simulation of the theory defined on a 4-dimensional Euclidean lattice.  Parton distribution functions (PDFs) encode important information on the internal structure of hadrons such as how the spin and momentum are distributed among the quarks and gluons.
However, these key quantities are presently only determined from global fits
to experimental data mostly for the proton.  Lattice QCD is thus the natural candidate for a reliable
computation of PDFs. However, since they are defined on the light-cone it was thought until recently to be impossible to compute PDFs directly on the lattice.
A method to extract PDFs from lattice QCD  was proposed by Ji few years ago and is based on the large-momentum effective theory
(LaMET) \cite{Ji:2013dva,Ji:2014gla}. In this approach, instead of calculating the light-cone quark
correlations in a hadron, one computes matrix elements with quarks separated by a space-like distance in
a boosted hadron, which is feasible on the lattice. Fourier transforming these spatial matrix elements
yields --after a suitable non-perturbative renormalization-- 
the so-called quasi-PDFs that  have the same infrared physics as  the light-cone PDFs. Because of this property,
one can relate quasi-PDFs and PDFs using perturbation theory, through a procedure called  matching.
If the hadron is boosted with high enough momentum, the matching equations should correctly
bring the quasi-PDFs to the PDFs, up to $O(M^2/P^2_z, \Lambda^2_{QCD}/P^2_z)$ corrections, where $M$ and $P_z$ are the
mass and momentum of the boosted hadron. The hadron mass corrections are known to all orders in $M^2/P_z^2$, while
the higher twist corrections are still not  calculated. Besides quasi-PDFs, there are other approaches to accessthe $x$-dependence PDFs from lattice QCD\cite{Detmold:2005gg, 
	Braun:2007wv,Ma:2014jla,Ma:2017pxb,Radyushkin:2017cyf}. We refer the interested Reader to the recent 
review of Ref.~\cite{Cichy:2018mum}.

Since Ji's proposal, considerable progress has been made. In particular progress in  the non-perturbative renormalization of 
quasi-PDFs \cite{Alexandrou:2017huk,Green:2017xeu,Ishikawa:2017faj} and in the calculation of the one-loop matching coefficient~\cite{Izubuchi:2018srq,Alexandrou:2018pbm} has enabled the extraction of PDFs from quasi-PDFs for the proton and pion.
 Results on the proton unpolarized PDFs have been obtained using also simulations  with physical pion mass~\cite{Chen:2018xof,Alexandrou:2019lfo,Alexandrou:2018pbm}.
 In this work we present first results on  the $\Delta (1232)$ baryon, the lightest baryon resonance of a 
nucleon-pion system. Being a strongly decaying resonance its structure is not directly accessible experimentally. Thus lattice QCD can provide important information on its structure by evaluating the  $\Delta$ electromagnetic and axial form factors and the $\Delta$-nucleon transition form factors. 
 In this work we present first results on the PDFs of the $\Delta(1232)$ baryon. Such a study
 is important in order to test the conjecture made in Ref.~\cite{Ethier:2018efr} that spontaneous breaking of chiral symmetry should lead to a significant difference between
the proton and the $\Delta^+$ flavor asymmetry for the isovector distribution $\overline{d}(x) - \overline{u}(x)$. Due to the short lifetime of the $\Delta$, lattice QCD offers the exciting possibility to study the $\Delta$ unpolarized PDF in order to understand the mechanism behind the breaking of the flavor symmetry between
light quarks in the nucleon sea.

\section{Theoretical setup}
The unpolarized PDF, denoted by $q(x)$, is defined on the light cone as~\cite{Collins:2011zzd}
\begin{equation}
q(x)=\int_{-\infty}^{+\infty} \frac{d \xi^{-}}{4 \pi} e^{-i x P^{+} \xi^{-}}\left\langle h\left|\overline{\psi}\left(0\right) \gamma^{+} W\left(0,\xi^{-}\right) \psi(\xi^{-})\right| h\right\rangle
\end{equation}
where the light-cone vectors are taken as $\xi^{ \pm}=\left(\xi^{0} \pm \xi^{3}\right) / \sqrt{2}$. $W\left(0,\xi^{-}\right)=e^{-i g \int_{0}^{\xi^{-}} d \eta^{-} A^{+}\left(\eta^{-}\right)}$ is the Wilson line required for gauge invariance and $x$ is the momentum fraction
carried by the quarks in the hadron. The plus component of the  momentum $P^{+}$ is $\left(P^{0}+P^{3}\right) / \sqrt{2}$ and  $|h\rangle$ the hadron state of interest. Deep inelastic scattering is light-cone
dominated, meaning that $\xi^{2}=t^{2}-\vec{r}^{2}\approxeq 0$. However, in Euclidean space time, this reduces to a 
single point, which makes it impossible to compute PDFs in lattice QCD directly. In the large momentum effective theory, the PDF can be extracted from the quasi-PDF defined by
\begin{equation}
\tilde{q}\left(x, {P}_{z}, \mu\right)=\int_{-\infty}^{+\infty} \frac{d z}{4 \pi} e^{-i x P_{z} z}\langle h(P_z)|\overline{\psi}(0) \Gamma {W}(0,{z}) \psi(z)| h(P_z)\rangle,
\end{equation}
where $|h(P_z)\rangle$ is the boosted hadron state with finite momentum $P=\left(E, 0,0, P_{z}\right)$, and $W(0,z)$ is the Wilson line along the boosted direction. $\mu$ is the renormalization scale.The Dirac structure $\Gamma$ defines the type of PDF. For the unpolarized PDF, $\Gamma$ is chosen to be $\gamma^0$ to avoid operator mixing~\cite{Constantinou:2017sej}.
\par
Since the infrared physics is the same for quasi-PDFs and light-cone PDFs and the difference is only in 
the perturbative region \cite{Ji:2014gla},  the quasi-PDFs are related to light-cone PDFs through the matching equation
\begin{equation}
\tilde{q}\left(x, P_{z}, \mu\right)=\int_{-1}^{1} \frac{d y}{|y|} C\left(\frac{x}{y}, \frac{\mu}{P_{z}}\right) q(y, \mu)+O\left(\frac{M^{2}}{P_{z}^{2}}, \frac{\Lambda_{Q C D}^{2}}{P_{z}^{2}}\right),
\end{equation}where $q(y,\mu)$ is the light-cone PDF at the scale $\mu$, C is the matching kernel, 
which can be calculated perturbatively and has been evaluated to one-loop level. 
How large the momentum $P_{z}$ needs to be for the  validity of  LaMET
has to be tested. Exploratory studies have been performed showing a great potential for this approach, as e.g. in Ref.~ \cite{Alexandrou:2016jqi,Alexandrou:2019lfo,Chen:2016utp,Chen:2018xof}.

\section{Lattice details}
\par
In this study, we evaluate the isovector unpolarized quasi-PDFs, $\overline{u}(x) - \overline{d}(x)$ of $\Delta^+$. The matrix elements of interest are given by
\begin{equation}\label{equ:square}
h(P_z, z)=\left\langle h\left|\overline{\psi}(0) \gamma^{0} W(0,z) \psi(z)\right| h\right\rangle
\end{equation}
for a straight Wilson line $W$, with varying length from $z=0$ up to half of the spatial extension $L/2$.  
In order to extract the matrix element of Eq. (\ref{equ:square}) we construct ratios of suitable three-point functions and two-point functions, averaged over the gauge field ensemble.
The two-point and three-point functions are given by
\begin{align}
C_{\sigma\rho}^{2 \mathrm{pt}}(\mathcal{P},\mathbf{P}; t_s, 0) &=\mathcal{P}_{\alpha \beta} \sum_{\mathbf{x}} e^{-i \mathbf{P} \cdot \mathbf{x}}\left\langle 0\left|{\cal J}_{\sigma \alpha}(\mathbf{x}, t_s) \overline{{\cal J}}_{\rho\beta}(\mathbf{0}, 0)\right| 0\right\rangle \\ C^{3 \mathrm{pt}}_{\sigma0\rho}\left(\tilde{\mathcal{P}},\mathbf{P} ; t_{s}, \tau, 0\right) &=\tilde{\mathcal{P}}_{\alpha \beta} \sum_{\mathbf{x}, \mathbf{y}} e^{-i \mathbf{P} \cdot \mathbf{x}}\left\langle 0\left|{\cal J}_{\sigma\alpha}\left(\mathbf{x}, t_{s}\right) \mathcal{O}(\mathbf{y}, \tau ; z) \overline{{\cal J}}_{\rho\beta}(\mathbf{0}, 0)\right| 0\right\rangle
\end{align}
where $t_s$ is the time separation of the sink relative to the source, $\tau$ is the insertion time of the $\mathcal{O}$ operator, and ${\cal J}_{\sigma\alpha}(x)$ is the $\Delta^+$ interpolating operator,
\begin{equation}
{\cal J}_{\sigma\alpha}(x)=\frac{1}{\sqrt{3}} \epsilon^{a b c}\left[2\left(u^{a^{T}}(x) C \gamma_{\sigma} d^{b}(x)\right)u_{\alpha}^{c}(x)+\left(u^{a^{T}}(x) C \gamma_{\sigma} u^{b}(x)\right) d_{\alpha}^{c}(x) \right],
\end{equation}
with $C=\gamma^0\gamma^2$ being the charge conjugation matrix. 

We use $\mathcal{P}=\tilde{\mathcal{P}}=\frac{1+\gamma^{0}}{4}$, and sum over the space components of 
$\sigma$,$\rho$ when computing the ratio of the three to the two point functions:

\begin{equation}
h\left(\mathbf{P}, z\right)\stackrel{ \tau, t_s>>1}{=}\frac{\sum_{\sigma=1}^{3} C_{\sigma 0 \sigma}^{3pt}\left(\tilde{\mathcal{P}},\mathbf{P}; t_s,\tau,0\right)}{ \sum_{\sigma=1}^{3} C^{2pt}_{\sigma \sigma}(\mathcal{P},\mathbf{P};t_s,0)}\; .
\label{eq:matrixelement}
\end{equation}
In principle, 
the interpolating field also overlaps with the heavier spin-1/2 excitations. 
However, it is known form previous studies that the overlap with spin-1/2 state is small, and can be neglected.

\section{Results at $m_\pi=330$ MeV}
As a first benchmark computation, we
use a $24^3 \times 48$ twisted mass ensemble with $a=0.096$ fm generated by the Extended Twisted Mass (ETM) collaboration \cite{Alexandrou:2018egz}. The twisted mass parameter is $\mu=0.0053$, which corresponds to a pion mass of $m_\pi =330 $ MeV. Since this is a first computation of the $\Delta$ PDF
we are interested in exploring the feasibility of the calculation and, in particular, the momentum 
smearing technique to reduce the statistical errors due to a boosted $\Delta$. 

It is well known that the overlap with the ground state  is improved by smearing the quark fields of the interpolating function. We use Gaussian smearing on the quark fields accompanied with APE smearing on the gauge fields that enter the Gaussian smearing function. For stationary hadron states this provides  an efficient approach to improve ground state dominance. However, in the computation of matrix elements of quasi-PDFs, we need the overlap of a  boosted particle state with the 
interpolating field. Thus, we use momentum smearing~\cite{Bali:2016lva} which has proven to improve significantly the overlap of a  boosted nucleon state with the interpolating field. The momentum smearing modifies the standard Gaussian smearing  by a complex phase, given by
\setlength\abovedisplayskip{1pt}
\setlength\belowdisplayskip{1pt}
\begin{equation}
S_{mom}=\frac{1}{1+6\alpha} \Big( \psi(x)+\alpha \sum_j U_j(x) e^{-i\xi P \cdot j} \psi(x+\hat{j}) \Big),
\end{equation} 
where $U_j$ is the gauge link in the spatial j-direction. The momentum smearing parameter $\xi$ needs to be tuned in order to optimize the overlap with the boosted particle state.

\begin{figure}[H]
	\centering
	\subfigure{\includegraphics[width=0.45\textwidth]{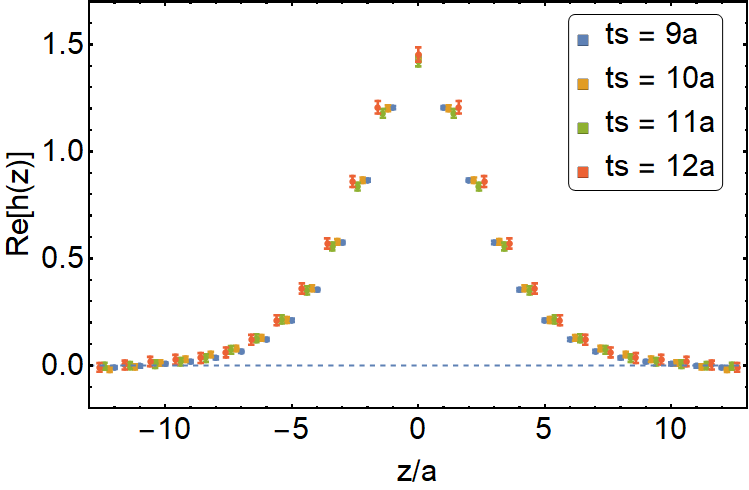}}
	\subfigure{\includegraphics[width=0.45\textwidth]{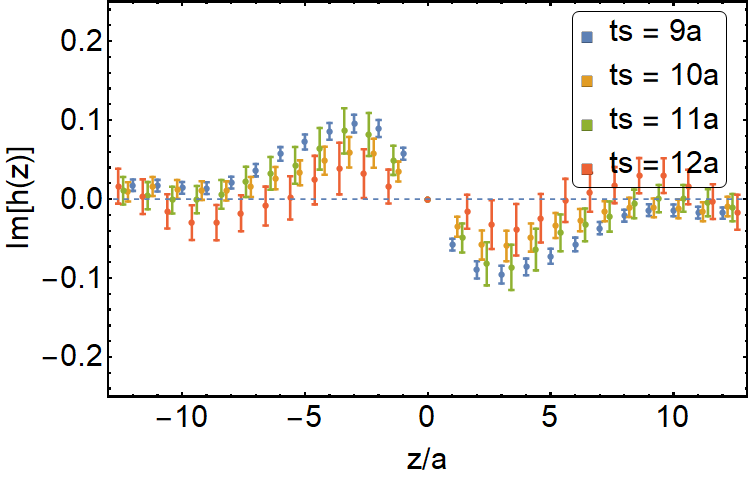}}
	\caption{The real (left) and imaginary (right) part of the ratio in 
eq.~(\ref{eq:matrixelement}) yielding the isovector unpolarized quasi-PDFs of the $\Delta$ with boosted momentum $P_z=2\pi/L$ for different values of the source-sink time separation as a function of $z/a$.}
	\label{differnet time separation}
\end{figure}

The necessity to boost the particle with relatively large momentum makes the excited state contamination more severe, since the spectrum gets denser. We use the source-sink time separations of {9a, 10a, 11a, 12a} 
corresponding to {0.86, 0.96, 1.05, 1.15} fm, in order to study the 
excited state contamination. In this preliminary study we employ 
momentum $P_z$ = $2\pi/L \approx$  0.54 GeV. In Fig.
\ref{differnet time separation}  we show the comparison of results among four different separations. The results for both the real part of the matrix element, for the four source-sink time separations are all consistent within our current precision. For the imaginary part there is a decreasing trend becoming consistent for source-sink time separation $t_s=10 a$. Therefore, we fix the $t_s$ to $10 a \approx 0.96$ fm throughout this work.

\section{Results at $m_\pi = 250$ MeV}
In this section we present results using  a $32^3 \times 64$ lattice with lattice spacing  $a=0.096$ fm and pion mass  $m_\pi = 250$~MeV. At this value of the pion mass  the ($\bar{d}-\bar{u}$) asymmetry is 
expected to be larger, as pointed out in Ref.~\cite{Ethier:2018efr}. All the results presented here 
are computed with a source-sink time separation of $10a$. We extract matrix elements for the first two momenta $\{2\pi/L,4\pi/L\}$, which correspond to $\{0.41,0.82\}$ GeV with $\{906,3600\}$ measurements.

\begin{figure}[h]
	\centering
	\subfigure{\includegraphics[width=0.45\textwidth]{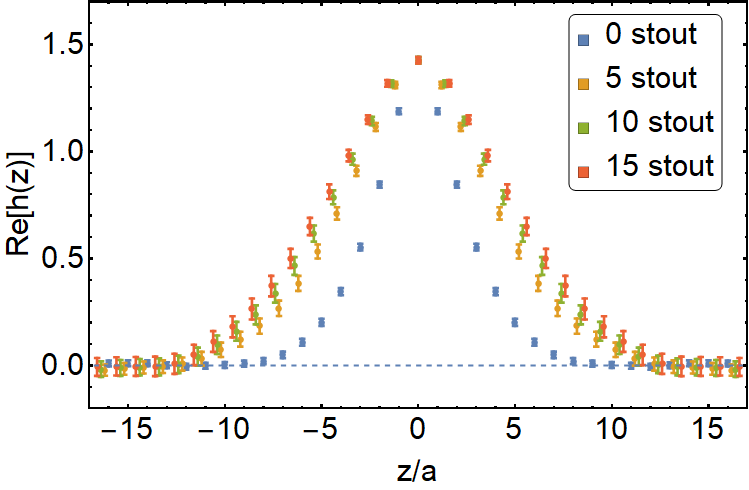}}
	\subfigure{\includegraphics[width=0.45\textwidth]{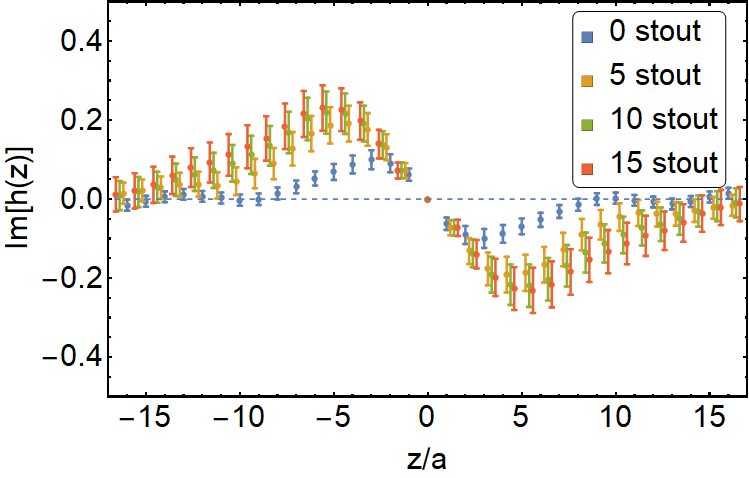}}
	\caption{The real (left) and imaginary (right) part of the ratio yielding the isovector unpolarized  quasi-PDF of the $\Delta$  for different smearing steps, for  momentum  $P_z=4\pi/L$.}
	\label{matrix element}
\end{figure}

In the computation of the matrix elements, we apply three-dimensional stout smearing to the gauge links that enter  in the Wilson line of the operator. This reduces the power divergence in the matrix elements and brings the necessary renormalization factors closer to tree-level value. 
In Fig.~\ref{matrix element}, we show the bare matrix elements for momentum $4\pi/L$ with different stout smearing steps. As can be seen, the stout smearing increases the value of matrix elements. The results for different stout smearing converge as the smearing steps increase.

\begin{figure}[h]
	\centering
	\subfigure{\includegraphics[width=0.45\textwidth]{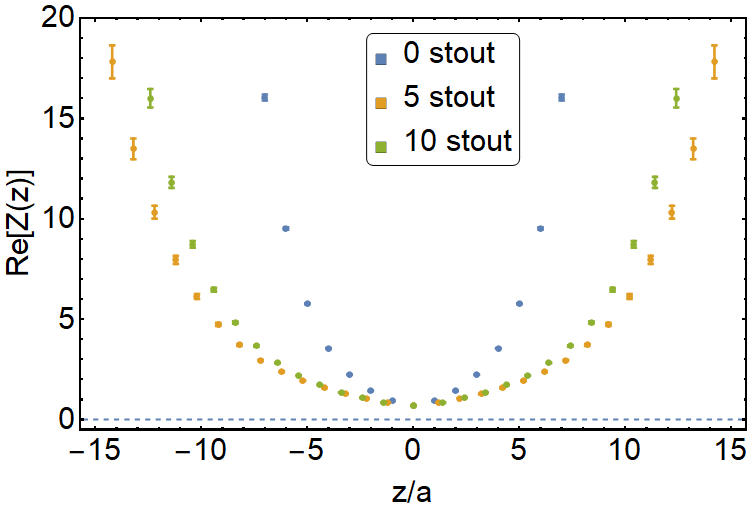}}
	\subfigure{\includegraphics[width=0.45\textwidth]{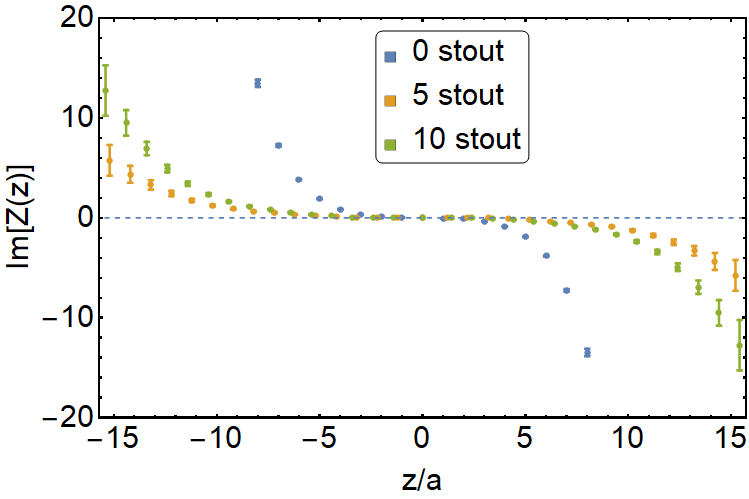}}
	\caption{Real part (left) and imaginary (right) parts of the renormalization factor for different stout smearing steps.}
	\label{renormalization factor}
\end{figure}

\begin{figure}[h]
	\centering
	\subfigure{\includegraphics[width=0.45\textwidth]{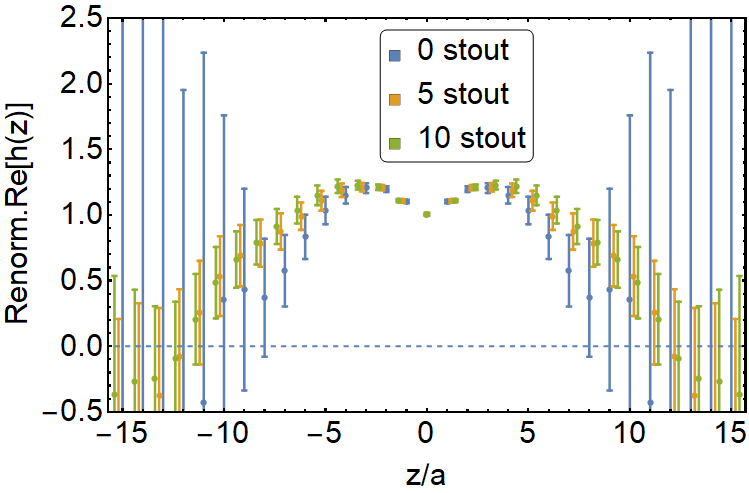}}
	\subfigure{\includegraphics[width=0.45\textwidth]{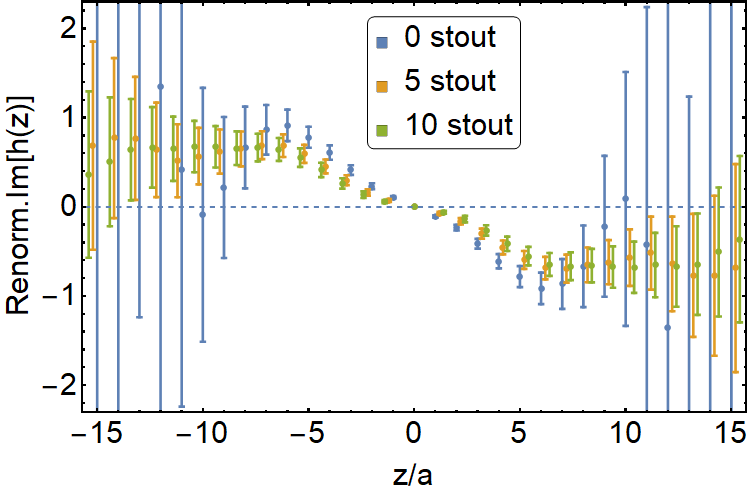}}
	\caption{Renormalized matrix of quasi-PDF with boosted momentum $4\pi/L$}
	\label{renormalization matrix}
\end{figure}

For obtaining physical results, the operator must be renormalized to eliminate divergences. We adopt a non-perturbative RI-MOM renormalization scheme~\cite{Martinelli:1994ty,Alexandrou:2017huk}. The renormalization factors $Z$ are converted to the $M\overline{MS}$ scheme~\cite{Alexandrou:2019lfo} and evolved to $\mu=2$ GeV. The final values of the  $Z$-factors are extracted by chirally extrapolating to zero pion mass.  The $Z$-factors are shown in Fig.\ref{renormalization factor} for different stout smearing steps.
The renormalized matrix elements are shown in Fig.\ref{renormalization matrix} for momentum $P_z={4\pi}/{L} \approx 0.82$ GeV. As expected, the  renormalized matrix elements for different stout smearing steps are consistent, with stout smearing  clearly 
reducing the errors in the renormalization matrix elements. The agreement between different smearing steps verifies the effectiveness of our renormalization procedure.
\section{Summary and outlook}
In this study, we perform a first and exploratory study of the renormalized matrix 
elements for the unpolarized isovector PDF of  the $\Delta^+$. 
The momentum boosts used are 0.41 GeV and 0.82 GeV for the 
$m_\pi=250$ MeV ensemble, and 0.54 GeV for the $m_\pi=330$ MeV 
ensemble. These values are rather small and the next step is 
therefore to boost the $ \Delta^+$ to higher momentum. In particular, we aim at reaching 
values of the momentum similar to those used in our studies of the nucleons quasi-PDFs.
This will enable us to  compute the ratio $(\overline{d}(x) - \overline{u}(x))^{\Delta^+}/(\overline{d}(x) - \overline{u}(x))^{p^+}$ in order to check the asymmetry enhancement prediction.
 If the light quark asymmetry in the nucleon sea has its origin in the spontaneous breaking of chiral symmetry, this ratio should be as large as 2 for $x\approxeq 0.1$. This would provide  a striking 
explanation of the physical mechanism responsible for $\overline{d}(x) > \overline{u}(x)$ in the nucleon.

\noindent
\textbf{Acknowledgments:} We would like to thank all members of Extended Twisted Mass Collaboration 
for their constant and pleasant collaboration. Y.C., X.F., C.L., Y.L. and S.X. were supported in part by NSFC of China under Grant No. 11775002 and No. 2015CB856702. It is also supported in part by DFG grant no. TRR~110 and NSFC grant No. 11621131001. This work has received funding from the European Union's Horizon 2020
research and innovation programme under the Marie Sk\l{}odowska-Curie grant agreement
No 642069 (HPC-LEAP). 
K.C.\ was supported by National Science Centre (Poland) grant SONATA
BIS no.\ 2016/22/E/ST2/00013. F.S.\ was funded by DFG project number 392578569.
M.C. acknowledges financial support by the U.S. Department of Energy, Office of Nuclear Physics, within
the framework of the TMD Topical Collaboration, as well as, by the National Science Foundation
under Grant No.\ PHY-1714407. This research used resources of TianHe-3 (prototype) at
Chinese National Supercomputer Center in Tianjin, and
JUWELS at  the Juelich Supercomputing Centre,  under project id ECY00.

\bibliography{reference}

\bibliographystyle{h-physrev}

\end{document}